\documentclass[12pt,a4paper]{article}
\usepackage{epsfig}
\begin{document}
\title{ The rare top quark decays $t\rightarrow cV$ in
the topcolor-assisted technicolor model}

\author{Gongru Lu$^{(a,b)}$, Furong  Yin$^{(b)}$, Xuelei Wang $^{(a,b)}$,
Lingde Wan $^{(b)}$   \\
 {\small a: CCAST (World Laboratory) P.O. BOX 8730. B.J. 100080 P.R. China}\\
 {\small b: College of Physics and Information Engineering,}\\
\small{Henan Normal University, Xinxiang  453002. P.R.China}
\thanks{This work is supported by the National Natural Science Foundation
of China(No.10175017), the Excellent Youth Foundation of Henan
Scientific Committee(No.02120000300), and the Henan Innovation
Project for University Prominent Research Talents(2002KYCX009).}
\thanks{E-mail:wangxuelei@263.net}}
\maketitle
\begin{abstract}
\hspace{5mm} We consider the rare top quark decays in the
framework of topcolor-assisted technicolor (TC2) model. We find
that the contributions of top-pions and top-Higgs predicted by the
TC2 model can enhance the SM branching ratios by as much as 6-9
orders of magnitude. i.e., in the most case, the orders of
magnitude of branching ratios are $Br(t\rightarrow c g )\sim
10^{-5}$, $Br(t\rightarrow c Z )\sim 10^{-5}$, $Br(t\rightarrow c
\gamma )\sim 10^{-7}$. With the reasonable values of the
parameters in TC2 model, such rare top quark decays may be
testable in the future experiments. So, rare top quark decays
provide us a unique way to test TC2 model.
\end {abstract}
\vspace{1.0cm} \noindent
 {\bf PACS number(s)}: 12.60Nz, 14.80.Mz,.15.LK, 14.65.Ha

\newpage
\begin{center}
{\bf I Introduction}
\end{center}

It is widely believed that the top quark, which with a mass of the
order of the electroweak scale, plays an important role in
particle physics. Its unusually large mass makes it more sensitive
to certain types of flavor-changing (FC) interactions.

In the standard model (SM), due to the GIM mechanism, the rare top
quark decays $t\rightarrow cV$ ($V=Z, \gamma, g$) are very
small\cite{y1}, far below the feasible experimental possibilites
at the future colliders(LHC or LC)\cite{y2}. In some new physics
models beyond the standard model(SM), the decay widths of the rare
top quark decays $t\rightarrow cV$ may be significantly enhanced
because of the appearance of large flavor changing couplings at
the tree-level. Various rare top quark decays have been
extensively studied in the SM \cite{y1}, the multi Higgs doublets
models(MHDM)\cite{y3}\cite{y4}\cite{y5}, the technicolor
models\cite{y6}\cite{y11}, the MSSM
models\cite{y7}\cite{y8}\cite{y9}, and other new physics models.
They have shown that, with reasonable values for the parameters,
the branching ratios $Br(t\rightarrow cV)$ could be within the
observable threshold of future experiments.

The topcolor-assisted technicolor(TC2) model\cite{y10} connects
the top quark with the electroweak symmetry breaking (EWSB). In
this model, the topcolor interactions make small contributions to
the EWSB, and give rise to the main part of the top quark mass
$(1-\epsilon)m_{t}$ with a model dependent parameter
$0.03\leq\epsilon\leq 0.1$. The technicolor (TC) interactions play
a main role in the breaking of the electroweak gauge symmetry. The
extend technicolor(ETC) interactions give rise to the masses of
the ordinary fermions including a very small portion of the top
quark mass $\epsilon m_{t}$. This kind of model predicts three
top-pions ($\Pi_{t}^{0}, \Pi_{t}^{\pm}$) and one top-Higgs
($h_{t}$) with large Yukawa couplings to the third generation.
These new particles can be regarded as a typical feature of the
TC2 model. Thus, studying the possible signature of these
particles and their contributions to some processes at high energy
colliders is a good method of testing the TC2 model. There have
been many publications related to this field
\cite{y11}\cite{y12}\cite{y13}. Another feature of the TC2 model
is the existence of large flavor-changing couplings. For TC2
models,  topcolor interactions are non-universal and therefore
does not posses a GIM mechanism, which results in a new
flavor-changing coupling vertices when one writes the interactions
in the quark mass eigen-basis. Thus, the top-pions and top-Higgs
predicted by this kind of models have large Yukawa couplings to
the third generation and can induce the new flavor-changing
couplings. Such flavor-changing couplings would give contributions
to the rare decays $t \rightarrow cV$. Because the rare top quark
decays $t\rightarrow cV$ can hardly be detected in the SM, any
observation of rare top quark decays would be an unambiguous
signal of new physics. So, the study of the rare top quark decays
within the framework of the TC2 model would be a feasible method
to test the TC2 model. Ref.\cite{y11} has considered the
contributions of these particles to the rare top quark decay
$t\rightarrow cg$. However Ref.\cite{y11} only considered the
contributions of neutral top-pion $\Pi_{t}^{0}$ and did not
consider the contributions of the charged top-pions
$\Pi_{t}^{\pm}$. In this paper, we systematically calculate the
contributions of the top-pions ($\Pi_{t}^{0}, \Pi_{t}^{\pm}$) and
top-Higgs ($h_{t}$) to the rare top quark decays $t\rightarrow cV$
in the TC2 model, and find that the TC2 model can significantly
enhanced the rare top quark decays $t\rightarrow cV$, and may
approach the detectability threshold of the future experiments.

\begin{center}
{\bf II The rare top quark decays $t\rightarrow cV$ in the TC2
model}
\end{center}

The TC2 model predicts the existence of the top-pions
$\Pi_{t}^{0}, \Pi_{t}^{\pm}$, top-pions would give the new flavor
changing couplings at tree-level. The relevant interactions of
these top-pions with the b, t and c quarks can be written as
\cite{y10}\cite{y12}:
\begin{eqnarray}
\nonumber
\frac{m_{t}}{\sqrt{2}F_{t}}\frac{\sqrt{\upsilon_{\omega}^{2}-
F_{t}^{2}}}{\upsilon_{\omega}}\hspace{-0.5cm}
&&[iK_{UR}^{tt}K_{UL}^{tt^{*}}\overline{t_{L}}t_{R}\Pi_{t}^{0}+
\sqrt{2}K_{UR}^{tt}K_{DL}^{bb^{*}}\overline{b_{L}}t_{R}\Pi_{t}^{-}+
iK_{UR}^{tc}K_{UL}^{tt^{*}}\overline{t_{L}}c_{R}\Pi_{t}^{0}
\\ && +\sqrt{2}K_{UR}^{tc}K_{DL}^{bb^{*}}\overline{b_{L}}c_{R}\Pi_{t}^{-}
+h.c.]
\end{eqnarray}
where $\upsilon_{\omega}=\upsilon/\sqrt{2}\approx 174GeV$, $F_{t}$
is the decay constant of the top-pions. $K_{UL}^{ij}$ are the
matrix elements of the unitary matrix $K_{UL}$ from which the
Cabibbo-Kobayashi-Maskawa (CKM) matrix can be derived as
$V=K_{UL}^{-1}K_{DL}$, and $K_{UR}^{ij}$ are the matrix elements
of the right-handed rotation matrix $K_{UR}$. Their values can be
written as:
\begin{eqnarray}
K_{UL}^{tt}=K_{DL}^{bb}=1,\hspace{0.5cm}
K_{UR}^{tt}=1-\epsilon,\hspace{0.5cm} K_{UR}^{tc}\leq
\sqrt{2\epsilon-\epsilon^{2}}
\end{eqnarray}
In the following calculation, we take
$K_{UR}^{tc}=\sqrt{2\epsilon-\epsilon^{2}}$ and take $\epsilon$ as
a free parameter.

The TC2 model also predicts a CP-even scalar $h_{t}$, called
top-Higgs \cite{y12}, which is a $\overline{t}t$ bound and
analogous to the $\sigma$ particle in low energy QCD. Its
couplings to quarks are similar to that of the neutral top-pion
except that the neutral top-pion is CP-odd. All the Feynman rules
of top-pions and top-Higgs relevant to $t \rightarrow cV$ are
shown in Appendix One.

The above large Yukawa couplings will effect the rare top quark
decays $t\rightarrow cV$. The relevant Feynman diagrams for the
contributions of the top-pions and top-Higgs to the rare top quark
decays $t\rightarrow cV$ are shown in Fig.1.
\begin{figure}[]
\vspace{-0.2cm}
   \begin{center}
\epsfig{file=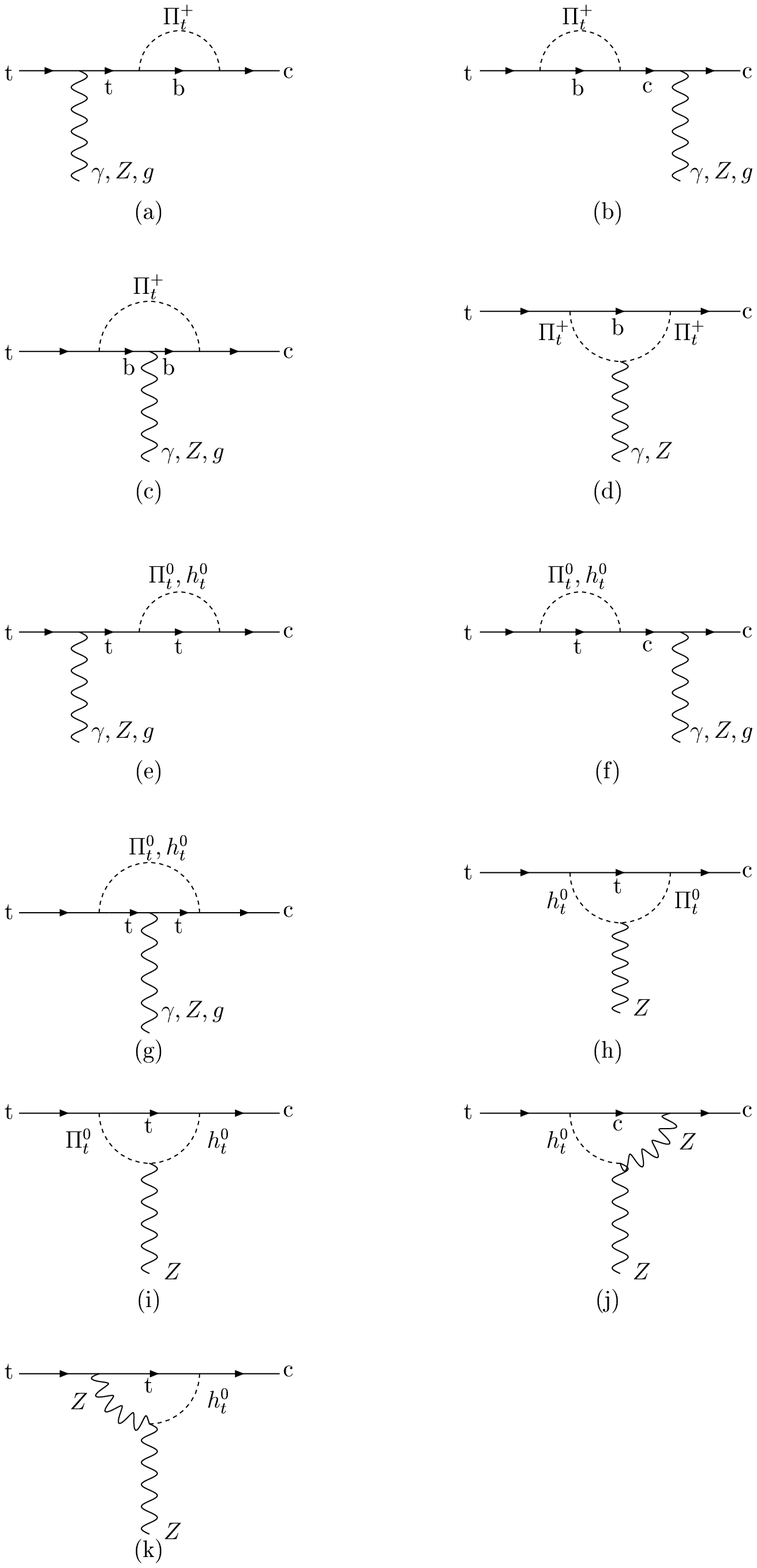, width=500pt,height=600pt}\vspace{-1cm}
 \caption {The
Feynman diagrams for the contributions of the
 top-pions($\Pi_{t}^{0}$, $\Pi_{t}^{\pm}$) and top-Higgs($h_{t}$) to the rare top quark
 decays $t\rightarrow cV$.}
 \label{fig.1}

 \end{center}
    \end{figure}
Using Eq.[1] and other
relevant Feynman rules, we obtain the relative amplitudes of the
rare top quark decays $t\rightarrow cV$:
\begin{eqnarray}
M_{V}=\overline{u_{c}} L (F_{V1} \gamma^{\mu}+F_{V2}
p_{t}^{\mu}+F_{V3} p_{c}^{\mu}) u_{t} \varepsilon_{\mu}(\lambda)
\end{eqnarray}
where $L=(1-\gamma_{5})/2$ is the left-handed projector, the
expressions of $F_{Vi}$($V=Z, \gamma, g$, $i=1,2,3$) in Eq.[3] can
be obtained by a straightforward calculations of the diagrams
shown in Fig.1.
 Because of $m_{t}>>m_{c}(m_{b}$), for the sake of simplicity, we
have neglected the terms proportional to $m_{c}, m_{b}$ in Eq.[3].
It can be seen that each diagram actually contain ultraviolet
divergences. Because there are no corresponding tree-level terms
to absorb these divergences, all the ultraviolet divergences
cancel in the effective vertex. Then, the widths of the rare top
quark decays contributed by top-pions and top-Higgs can be written
as:
\begin{eqnarray}
\nonumber
 \Gamma(t\rightarrow cZ)&=&\frac{1}{16\pi
m_{t}}(1-\frac{M_{Z}^{2}}{m_{t}^{2}})\frac{1}{8M_{Z}^{2}}
[F_{Z1}^{2}(4m_{t}^{2}M_{Z}^{2}-8M_{Z}^{4}+4m_{t}^{4})\\&&
\nonumber+
F_{Z2}^{2}(-3m_{t}^{4}M_{Z}^{2}+m_{t}^{6}+3m_{t}^{2}M_{Z}^{4}-M_{Z}^{6})
+F_{Z3}^{2}(m_{t}^{2}-M_{Z}^{2})^{3}\nonumber
\\&& + (F_{Z1}\cdot F_{Z2}^{*}+F_{Z2}\cdot
F_{Z1}^{*})(-4m_{t}^{3}M_{Z}^{2}+2m_{t}^{5}+2M_{Z}^{4}m_{t})
\nonumber
\\&&+(F_{Z2}\cdot F_{Z3}^{*}+F_{Z3}\cdot
F_{Z2}^{*})(-3m_{t}^{4}M_{Z}^{2}+m_{t}^{6}+3m_{t}^{2}M_{Z}^{4}-M_{Z}^{6})
\nonumber
\\&&+2(F_{Z1}\cdot F_{Z3}^{*}+F_{Z3}\cdot
F_{Z1}^{*})(m_{t}^{2}-M_{Z}^{2})^{2}m_{t}]
\end{eqnarray}
\begin{eqnarray}
\nonumber
 \Gamma(t\rightarrow c\gamma)&=&\frac{1}{16\pi m_{t}}
 [F_{\gamma 1}^{2}m_{t}^{2}-\frac{1}{2}F_{\gamma 2}^{2}m_{t}^{4}
 -\frac{1}{2}(F_{\gamma 1}F_{\gamma 2}^{*}+F_{\gamma 2}F_{\gamma
 1}^{*})m_{t}^{3}\nonumber\\&&
 -\frac{1}{4}(F_{\gamma 2}F_{\gamma 3}^{*}+F_{\gamma 3}F_{\gamma
 2}^{*})m_{t}^{4}]
\end{eqnarray}
\begin{eqnarray}
\nonumber
 \Gamma(t\rightarrow cg)&=&\frac{1}{16\pi m_{t}}
 [F_{g 1}^{2}m_{t}^{2}-\frac{1}{2}F_{g 2}^{2}m_{t}^{4}
 -\frac{1}{2}(F_{g 1}F_{g 2}^{*}+F_{g 2}F_{g 1}^{*})m_{t}^{3}
 \nonumber\\&&
 -\frac{1}{4}(F_{g 2}F_{g 3}^{*}+F_{g 3}F_{g 2}^{*})m_{t}^{4}]
\end{eqnarray}
where $m_{t}$ and $M_{Z}$ denote the masses of top quark and $Z$
boson, respectively. The explicit expressions of the form factors
$F_{\gamma i},F_{Z i},F_{gi}$ are given in Appendix Two.

\begin{center}
{\bf III The numerical results and conclusions }
\end{center}

According to the above calculations, we can give the numerical
results of the branching ratio of $t\rightarrow cV$ contributed by
$\Pi_t$ and $h^0_t$. In this paper, we adopt the branching ratios
$Br(t\rightarrow cV)$ defined as\cite{y1}:
\begin{eqnarray}
Br(t\rightarrow cV)=\frac{\Gamma(t\rightarrow
cV)}{\Gamma(t\rightarrow W^{+} b)}
\end{eqnarray}

Before numerical calculations, we need to specify the parameters
involved. We take $m_{t}=175$ GeV, $M_{Z}=91.18$ GeV,
$s_{W}^{2}=sin^{2}\theta_{W}=0.23$, $\alpha_{e}=1/128.9$ and
$\alpha_{s}=0.118$. Now, there are still four free parameters:
$\epsilon$, $m_{\Pi_{t}^{0}}$, $m_{\Pi_{t}^{\pm}}, m_{h_{t}}$.
$\epsilon$ is a model dependent parameter and we take it in the
range of $0.03\sim 0.1$, $m_{\Pi_{t}^{0}}$, $m_{\Pi_{t}^{\pm}},
m_{h_{t}}$ denote the masses of neural top-pion $\Pi_{t}^{0}$,
charged top-pion $\Pi_{t}^{\pm}$ and top-Higgs ($h_{t}$),
respectively. Due to the split of the $m_{\Pi_{t}^{0}}$ and
$m_{\Pi_{t}^{\pm}}$ only come from the electroweak interactions,
the different of $m_{\Pi_{t}^{0}}$ and $m_{\Pi_{t}^{\pm}}$ is very
small and can be ignored. Here, we take
$m_{\Pi_{t}^{0}}=m_{\Pi_{t}^{\pm}}=m_{\Pi_{t}}$. Ref.\cite{y10}
have estimated the mass of top-pions, the results show that the
$m_{\Pi_{t}}$ is allowed to be in the region of a few hundred GeV
depending on the models. Estimating the contributions of top-pions
to the rare top quark decays $t\rightarrow cV$, we take the mass
of top-pion to vary in the range of 200 GeV $\sim$ 500 GeV in this
paper. The mass of $h_{t}$ can be estimated in the
Nambu-Jona-Lasinio (NJL) model in the large $N_{c}$ approximation
and is found to be of the order of $m_{h_{t}}\approx 2m_{t}$
\cite{y12}. This estimation is rather crude and the masses well
below the $\overline{t}t$ threshold are quite possible and occur
in a variety of cases \cite{y15}. As the branching ratios are
proportional to $(2\epsilon-\epsilon^{2})(1-\epsilon)^{2}$, to
cancel the influence of $\epsilon$ on the branching ratio, we
summarized the final numerical results of $\frac{Br(t\rightarrow
cV)}{(2\epsilon-\epsilon^{2})(1-\epsilon)^{2}}$ in Figs. 2-4.
\begin{figure}[]
\vspace{-0.5cm}
\begin{center} \epsfig{file=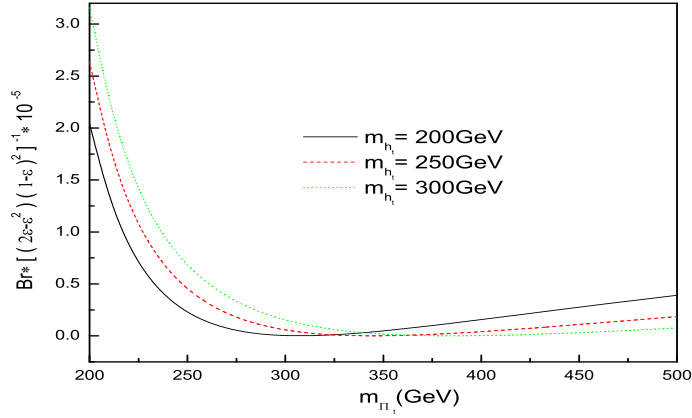, width=292pt,height=180pt}
 \vspace{-0.6cm}
 \caption {The branching ratio $\frac{Br(t\rightarrow
c\gamma)}{(2\epsilon-\epsilon^{2})(1-\epsilon)^{2}}$ as a function
of top-pion mass $m_{\Pi_{t}}$ for the mass of top-Higgs
$m_{h_{t}}=200$ GeV(solid line), $m_{h_{t}}=250$ GeV(dashed line),
$m_{h_{t}}=300$ GeV(dotted line), respectively.}
 \label{fig.2}
\end{center}
\end{figure}
\begin{figure}[]
\vspace{-1.2cm}
\begin{center}
 \epsfig{file=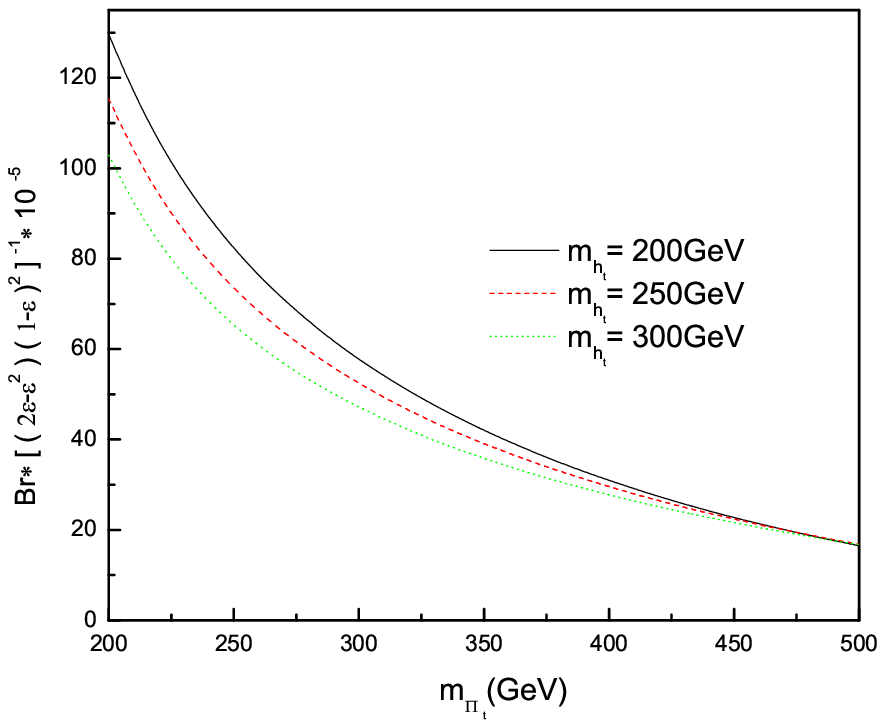, width=292pt,height=180pt}
 \vspace{-0.5cm}
 \caption {The same as Fig.2 but for the process of $t\rightarrow
c Z$} \label{fig.3}
\end{center}
\end{figure}
\begin{figure}[]
\begin{center}
 \vspace{-0.2cm}
 \epsfig{file=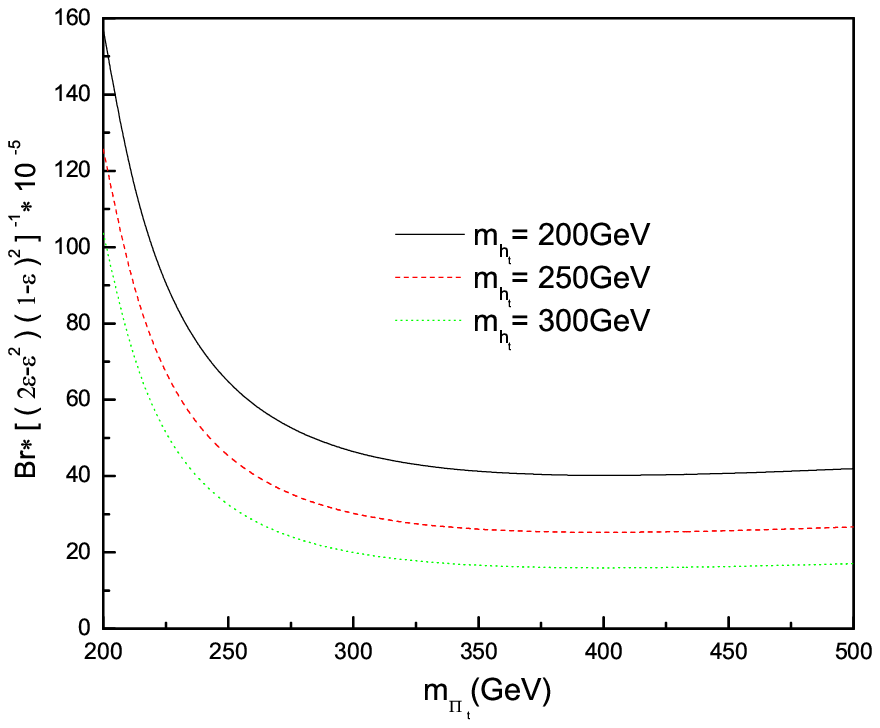,width=292pt,height=180pt}
 \caption {The same as Fig.2 but for the process of $t\rightarrow
c g$} \label{fig.4}
\end{center}
\end{figure}

Fig.2-4 are the plots of the $\frac{Br(t\rightarrow c V
)}{(2\epsilon-\epsilon^{2})(1-\epsilon)^{2}}$ versus
$m_{\Pi_{t}}(200$ GeV $\sim 500$ GeV) for $m_{h_{t}}=200$ GeV, 250
GeV, 300 GeV, respectively. We can see that, the branching ratio
of $t\rightarrow c \gamma$ is two order smaller than that of
$t\rightarrow c Z$ and $t\rightarrow c g$. The $Br(t\rightarrow c
\gamma)$ decreases as $m_{\Pi_{t}}$ increase and $m_{h_{t}}$
decrease for small $m_{\Pi_{t}}$, but for large $m_{\Pi_{t}}$, it
increases with $m_{\Pi_{t}}$ increasing and $m_{h_{t}}$
decreasing. The $Br(t\rightarrow c Z )$ are very sensitive to
top-pions mass and decreases with $m_{\Pi_{t}}$ and $m_{h_{t}}$
increasing. But for very large $m_{\Pi_{t}}$, the branching ratio
of $t\rightarrow c Z$ hardly changes with the $m_{h_{t}}$. As for
$t\rightarrow c g$, the branching ratio decrease very sharply as
$m_{\Pi_{t}}$ increase for small $m_{\Pi_{t}}$. In the most case,
the orders of magnitude of branching ratios are $Br(t\rightarrow c
g )\sim 10^{-5}$, $Br(t\rightarrow c Z )\sim 10^{-5}$,
$Br(t\rightarrow c \gamma )\sim 10^{-7}$.

Comparing with the theoretical predictions in the other models, we
list the maximum levels of $Br(t\rightarrow cV)$ predicted by the
SM \cite{y1}, the MSSM \cite{y5} and the TC2 model as follows:
\begin{center}
\doublerulesep 0.8pt \tabcolsep 0.2in
\begin{tabular}{||c|c|c|c||}\hline\hline
 & SM & MSSM & TC2\\ \hline
$Br(t\rightarrow cZ)$ & $O(10^{-13})$ & $O(10^{-7})$ & $O(10^{-4})$\\
\hline
$Br(t\rightarrow c\gamma)$ & $O(10^{-13})$ & $O(10^{-7})$ & $O(10^{-6})$\\
\hline $Br(t\rightarrow cg)$ & $O(10^{-11})$ & $O(10^{-4})$ &
$O(10^{-4})$
\\ \hline\hline
\end{tabular}
\end{center}
\begin{center}
\null\noindent ~~{\bf Tbale 1:}Theoretical predictions for
branching ratios of the rare top quark decays $t\rightarrow cV$.
\end{center}
It is shown that the branching ratios of $t\rightarrow cV$ in TC2
model are significant large than that in SM and MSSM. The
contributions of $\Pi_t$ and $h_t$ can enhance the SM branching
ratios by as much as 6-9 orders of magnitude. On the other hand,
$Br(t\rightarrow cZ)$ predicted by TC2 model is about 3 orders of
magnitude larger than that predicted by MSSM. So, the mode of
$t\rightarrow cZ$ is especially important for us to distinguish
TC2 from MSSM.

To assess the discovery reach of the rare top quark decays in the
future high energy colliders, Ref.\cite{y16} has roughly estimated
the following sensitivities for $100fb^{-1}$ of integrated
luminosity:
\begin{eqnarray}
LHC:    Br(t\rightarrow cV)\geq 5\times 10^{-5},\\
LC:     Br(t\rightarrow cV)\geq 5\times 10^{-4},\\
TEV33:  Br(t\rightarrow cV)\geq 5\times 10^{-3}.
\end{eqnarray}
Comparing the theoretical predictions in TC2 model with the
sensitivities of future high luminosity colliders(LHC,LC,TEV33),
we can conclude that TC2 model can enhance the branching ratios
$Br(t\rightarrow cV)$ to be within the observable threshold of
future experiments, especially for $t\rightarrow cZ$. LHC seems to
be the most suitable collider where to test rare top quark decays.
The LC is limited by statistics but in compensation every
collected event is clear-cut. So, this machine could eventually be
of much help, especially for high luminosity.

In conclusion, we have calculated the rare top quark decays
$t\rightarrow cV$ in the TC2 model. We find that the contributions
arising from $\Pi_{t}$ and $h_{t}$ predicted by the TC2 model
indeed significantly enhance the branching ratios of the rare top
quark decays. The channels $t\rightarrow cZ$ and $t\rightarrow cg$
are found to have the larger branching ratios, which can reach
$10^{-4}$ for the favorable parameter values and may be detectable
in the future high energy colliders.  Therefore, the rare top
quark decays provide us a unique way to test TC2. Otherwise,
$Br(t\rightarrow cZ)$ predicted by TC2 model is much larger than
that predicted by MSSM. So, we can distinguish TC2 from MSSM via
$t\rightarrow cZ$ mode.
\newpage
\begin{center}
{\bf Acknowledgments}
\end{center}

We thank J J, Cao for the valuable discussions.

\appendix
\setcounter{equation}{0}
\begin{center}
{\bf Appendix One: The feynman rules needed in the calculations}
\end{center}

Based on the effective Yukawa coupings to ordinary fermions of the
top-pions and top-Higgs in the TC2 model, we can write down the
relevant Feynman rules used in this paper:
\begin{eqnarray}
\Pi_{t}^{0}\overline{t}t:&&
-\frac{m_{t}}{\sqrt{2}F_{t}}\frac{\sqrt{\upsilon_{\omega}^{2}-F_{t}^{2}}}{\upsilon_{\omega}}
(1-\epsilon)\gamma_{5} \\
 \Pi_{t}^{0}\overline{t}c:&&
\frac{m_{t}}{\sqrt{2}F_{t}}\frac{\sqrt{\upsilon_{\omega}^{2}-F_{t}^{2}}}{\upsilon_{\omega}}
\frac{1-\gamma_{5}}{2}\sqrt{2\epsilon-\epsilon^{2}}\\
\Pi_{t}^{+}\overline{t}b:&&
i\sqrt{2}\frac{m_{t}}{\sqrt{2}F_{t}}\frac{\sqrt{\upsilon_{\omega}^{2}-F_{t}^{2}}}
{\upsilon_{\omega}} \frac{1+\gamma_{5}}{2}(1-\epsilon)\\
\Pi_{t}^{0}\overline{b}c: &&
i\sqrt{2}\frac{m_{t}}{\sqrt{2}F_{t}}\frac{\sqrt{\upsilon_{\omega}^{2}-F_{t}^{2}}}
{\upsilon_{\omega}}\frac{1-\gamma_{5}}{2}\sqrt{2\epsilon-\epsilon^{2}}\\
h_{t}^{0}\overline{t}t:&&
\frac{im_{t}}{\sqrt{2}F_{t}}\frac{\sqrt{\upsilon_{\omega}^{2}-F_{t}^{2}}}{\upsilon_{\omega}}
(1-\epsilon)\\
 h_{t}^{0}\overline{t}c: &&
\frac{im_{t}}{\sqrt{2}F_{t}}\frac{\sqrt{\upsilon_{\omega}^{2}-F_{t}^{2}}}{\upsilon_{\omega}}
\frac{1-\gamma_{5}}{2}\sqrt{2\epsilon-\epsilon^{2}}\\
Zh_{t}^{0}\Pi_{t}^{0}:&&\frac{g}{2c_{W}}(p_{1}-p_{2})_{\mu}\\
ZZh_{t}^{0}:&&i\frac{F_{t}}{\upsilon_{\omega}}\frac{gM_{Z}}{c_{W}}g_{\mu\nu}
\end{eqnarray}
$g=\frac{e}{2c_{W}}$, $c_{W}=cos\theta_{W}$ is the Weinberg angle.
 \setcounter{equation}{0}
\begin{center}
{\bf Appendix Two: The explicit expressions of the form factors:
$F_{Vi}$ }
\end{center}
The explicit expressions of the form factors $F_{Vi}$ used in
(3)-(6) can be written as:
\begin{eqnarray}
F_{Z i}=k_{Z}\sum_{\alpha=a}^{i}F_{Zi}^{\alpha}+
k_{Z}^{\prime}\sum_{\beta=j}^{k}F_{Zi}^{\beta}; \hspace{0.2cm}
F_{\gamma i}=k_{\gamma}\sum_{\alpha=a}^{i}F_{\gamma
i}^{\alpha};\hspace{0.2cm}
 F_{g
i}=k_{g}\sum_{\alpha=a}^{i}F_{gi}^{\alpha}
\end{eqnarray}
Here, $\alpha=a,b,c,d,e,f,g,h,i$ and $\beta=j,k$ denote each
Feynman diagrams in Fig.1. $F_{Vi}^{\alpha}(V=\gamma, Z,g$
$i=1,2,3)$ are the contributions arising from corresponding
Feynman diagrams.
\begin{eqnarray}
F_{Z1}^{b}&=&\frac{8}{3}s_{W}^{2}(B_{0}^{b}+B_{1}^{b})\\\nonumber
F_{Z1}^{c}&=&2(1-\frac{2}{3}s_{W}^{2})[-m_{t}^{2}(C_{11}^{c}-C_{12}^{c}+C_{21}^{c}
+C_{22}^{c}-2C_{23}^{c})\\&&
+(m_{t}^{2}-M_{Z}^{2})(C_{22}^{c}-C_{23}^{c})-2C_{24}^{c}+\frac{1}{2}]\\
F_{Z2}^{c}&=&-4(1-\frac{2}{3}s_{W}^{2})m_{t}(C_{22}^{c}-C_{23}^{c})\\
F_{Z3}^{c}&=&4(1-\frac{2}{3}s_{W}^{2})m_{t}(C_{11}^{c}-C_{12}^{c}+C_{21}^{c}+
C_{22}^{c}-2C_{23}^{c})\\
F_{Z1}^{d}&=&4(1-2s_{W}^{2})C_{24}^{d}\\
F_{Z2}^{d}&=&-2(1-2s_{W}^{2})m_{t}(4C_{23}^{d}-2C_{22}^{d}-2C_{21}^{d}+C_{12}^{d}
-C_{11}^{d})\\
F_{Z3}^{d}&=&-2(1-2s_{W}^{2})m_{t}(2C_{22}^{d}-2C_{23}^{d}+C_{12}^{d}
-C_{11}^{d})\\
F_{Z1}^{e}&=&\frac{4}{3}s_{W}^{2}(B_{0}^{e}-B_{0}^{*e})\\
F_{Z1}^{f}&=&\frac{4}{3}s_{W}^{2}(B_{1}^{f}+B_{1}^{*f}+2B_{0}^{*f})\\\nonumber
F_{Z1}^{g}&=&(1-\frac{4}{3}s_{W}^{2})[m_{t}^{2}(C_{22}^{g}-2C_{23}^{g}+C_{21}^{g}
+C_{22}^{*g}-2C_{23}^{*g}+C_{21}^{*g})
\\\nonumber
&&+(2C_{24}^{g}+2C_{24}^{*g}-1)]-\frac{4}{3}s_{W}^{2}
m_{t}^{2}(C_{11}^{g}-C_{12}^{g}+C_{11}^{*g}-C_{12}^{*g})\\
&&-(1-\frac{4}{3}s_{W}^{2})(m_{t}^{2}-M_{Z}^{2})(C_{22}^{g}-
C_{23}^{g}+C_{22}^{*g}-C_{23}^{*g}) +
\frac{8}{3}s_{W}^{2}m_{t}^{2}C_{0}^{*g}\\
F_{Z2}^{g}&=&-2(1-\frac{4}{3}s_{W}^{2})m_{t}(C_{21}^{g}+C_{22}^{g}-2C_{23}^{g}
+C_{21}^{*g}+C_{22}^{*g}-
2C_{23}^{*g}-2C_{12}^{*g}-2C_{11}^{*g})\\
F_{Z3}^{g}&=&-2(1-\frac{4}{3}s_{W}^{2})m_{t}(C_{23}^{g}-C_{22}^{g}+
C_{23}^{*g}-C_{22}^{*g})-
\frac{8}{3}s_{W}^{2}m_{t}(C_{12}^{g}-C_{12}^{*g})\\
F_{Z1}^{h}&=&-2C_{24}^{h}\\
F_{Z2}^{h}&=&m_{t}(4C_{23}^{h}-2C_{22}^{h}-2C_{21}^{h}+C_{0}^{h}-C_{12}^{h}+C_{11}^{h})\\
F_{Z3}^{h}&=&m_{t}(2C_{22}^{h}-2C_{23}^{h}+3C_{12}^{h}-C_{11}^{h}+C_{0}^{h})\\
F_{Z1}^{i}&=&-2C_{24}^{i}\\
F_{Z2}^{i}&=&m_{t}(4C_{23}^{i}-2C_{22}^{i}-2C_{21}^{i}-C_{0}^{i}+3C_{12}^{i}-3C_{11}^{i})\\
F_{Z3}^{i}&=&m_{t}(2C_{22}^{i}-2C_{23}^{i}-C_{12}^{i}-C_{11}^{i}-C_{0}^{i})\\
F_{Z1}^{j}&=&\frac{4}{3}s_{W}^{2}m_t(C_{12}^{j}-C_{11}^{j})\\
F_{Z3}^{j}&=&-\frac{8}{3}s_{W}^{2}C_{12}^{j}\\
F_{Z1}^{k}&=&m_{t}[1-\frac{4}{3}s_{W}^{2})(C_{12}^{k}-C_{11}^{k})
+\frac{4}{3}s_{W}^{2}C_{0}^{k}]\\
F_{Z2}^{k}&=&-2(1-\frac{4}{3}s_{W}^{2})(C_{12}^{k}-C_{11}^{k})\\
F_{\gamma1}^{b}&=&-\frac{4}{3}(B_{0}^{b}+B_{1}^{b})\\
F_{\gamma1}^{c}&=&\frac{2}{3}[-m_{t}^{2}(C_{11}^{c}-C_{12}^{c}+C_{21}^{c}-C_{23}^{c})
-2C_{24}^{c}+\frac{1}{2}]\\
F_{\gamma2}^{c}&=&\frac{4}{3}m_{t}(C_{11}^{c}-C_{12}^{c}+C_{21}^{c}+
C_{22}^{c}-2C_{23}^{c})\\
F_{\gamma3}^{c}&=&-\frac{4}{3}m_{t}(C_{22}^{c}-C_{23}^{c})\\
F_{\gamma1}^{d}&=&4C_{24}^{d}\\
F_{\gamma2}^{d}&=&-2m_{t}(4C_{23}^{d}-2C_{22}^{d}-2C_{21}^{d}+C_{12}^{d}
-C_{11}^{d})\\
F_{\gamma3}^{d}&=&-2m_{t}(2C_{22}^{d}-2C_{23}^{d}+C_{12}^{d}
-C_{11}^{d})\\
F_{\gamma1}^{e}&=&-\frac{2}{3}(B_{0}^{e}-B_{0}^{*e})\\
F_{\gamma1}^{f}&=&-\frac{2}{3}(B_{1}^{f}+B_{1}^{*f}+2B_{0}^{*f})\\\nonumber
F_{\gamma1}^{g}&=&-\frac{2}{3}[-m_{t}^{2}(C_{21}^{g}-C_{23}^{g}+C_{11}^{g}-C_{12}^{g}
+C_{21}^{*g}-C_{23}^{*g}+C_{11}^{*g}-C_{12}^{*g})\\
&&+(-2C_{24}^{g}-2C_{24}^{*g}+1)]
-\frac{4}{3}m_{t}^{2}C_{0}^{*g}\\
F_{\gamma2}^{g}&=&-\frac{4}{3}m_{t}(C_{21}^{g}+C_{22}^{g}-2C_{23}^{g}
+2C_{11}^{*g}+C_{22}^{*g}-
2C_{23}^{*g}+C_{21}^{*g}-2C_{12}^{*g})\\
F_{\gamma3}^{g}&=&-\frac{4}{3}m_{t}[C_{23}^{g}-C_{22}^{g}-C_{12}^{g}
+C_{23}^{*g}-C_{22}^{*g}+C_{12}^{*g}]\\
F_{g1}^{b}&=&-2(B_{0}^{b}+B_{1}^{b})\\
F_{g1}^{c}&=&-2[-m_{t}^{2}(C_{11}^{c}-C_{12}^{c}+C_{21}^{c}-C_{23}^{c})
-2C_{24}^{c}+\frac{1}{2}]\\
F_{g2}^{c}&=&-4m_{t}(C_{11}^{c}-C_{12}^{c}+C_{21}^{c}+
C_{22}^{c}-2C_{23}^{c})\\
F_{g3}^{c}&=&4m_{t}(C_{22}^{c}-C_{23}^{c})\\
F_{g1}^{e}&=&-(B_{0}^{e}-B_{0}^{*e})\\
F_{g1}^{f}&=&-(B_{1}^{f}+B_{1}^{*f}+2B_{0}^{*f})\\\nonumber
F_{g1}^{g}&=&m_{t}^{2}(C_{21}^{g}-C_{23}^{g}+C_{11}^{g}-C_{12}^{g}
+C_{21}^{*g}-C_{23}^{*g}+C_{11}^{*g}-C_{12}^{*g})\\
&&-2C_{24}^{g}-2C_{24}^{*g}+1-2m_{t}^{2}C_{0}^{*g}\\
F_{g2}^{g}&=&-2m_{t}(C_{21}^{g}+C_{22}^{g}-2C_{23}^{g}+2C_{11}^{*g}+C_{22}^{*g}-
2C_{23}^{*g}+C_{21}^{*g}-2C_{12}^{*g})\\
F_{g3}^{g}&=&-2m_{t}[C_{23}^{g}-C_{22}^{g}+C_{23}^{*g}-C_{22}^{*g}+C_{12}^{g}-C_{12}^{*g}]
\end{eqnarray}
\begin{eqnarray}\nonumber
k_{Z}&=&-\frac{i}{16\pi^{2}}\frac{m_{t}^{2}}{2F_{t}^{2}}\frac{\upsilon_{\omega}^{2}-F_{t}^{2}}
{\upsilon_{\omega}^{2}}\sqrt{2\epsilon-\epsilon^{2}}(1-\epsilon)\frac{g}{2c_{W}}\\\nonumber
k_{Z}^{\prime}&=&-\frac{i}{32\pi^{2}}\frac{m_{t}M_Z}{\sqrt{2}\upsilon_{\omega}}
\frac{\sqrt{\upsilon_{\omega}^{2}-F_{t}^{2}}}{\upsilon_{\omega}}\frac{g^2}
{c_{W}^2}\\ \nonumber
k_{\gamma}&=&-\frac{i}{16\pi^{2}}\frac{m_{t}^{2}}{2F_{t}^{2}}
\frac{\upsilon_{\omega}^{2}-F_{t}^{2}}
{\upsilon_{\omega}^{2}}\sqrt{2\epsilon-\epsilon^{2}}(1-\epsilon)e\\
k_{g}&=&-\frac{i}{16\pi^{2}}\frac{m_{t}^{2}}{2F_{t}^{2}}\frac{\upsilon_{\omega}^{2}-F_{t}^{2}}
{\upsilon_{\omega}^{2}}\sqrt{2\epsilon-\epsilon^{2}}(1-\epsilon)g_{s}T^{\alpha}
\end{eqnarray}
\begin{eqnarray}\nonumber
B_{i}^{b}=B_{i}(-p_{t},m_{\Pi_{t}^{\pm}},m_{b});&&
B_{i}^{e}=B_{n}(-p_{c},m_{\Pi_{t}^{0}},m_{t});\\\nonumber
B_{i}^{*e}=B_{i}(-p_{c},m_{h_{t}},m_{t});&&
B_{i}^{f}=B_{i}(-p_{t},m_{\Pi_{t}^{0}},m_{t});\\\nonumber
B_{i}^{*f}=B_{i}(-p_{t},m_{h_{t}},m_{t});&&
C_{ij}^{c}=C_{ij}(-p_{t},p_{V},m_{\Pi_{t}^{\pm}},m_{b},m_{b});\\\nonumber
C_{ij}^{d}=C_{ij}(-p_{t},p_{V},m_{b},m_{\Pi_{t}^{\pm}},m_{\Pi_t^{\pm}});&&
C_{ij}^{g}=C_{ij}(-p_{t},p_{V},m_{\Pi_{t}^{0}},m_{t},m_{t});\\\nonumber
C_{ij}^{*g}=C_{ij}(-p_{t},p_{V},m_{h_{t}},m_{t},m_{t});&&
C_{ij}^{h}=C_{ij}(-p_{t},p_{V},m_{t},m_{h_{t}},m_{\Pi_{t}^{0}});\\\nonumber
C_{ij}^{i}=C_{ij}(-p_{t},p_{V},m_{t},m_{\Pi_{t}^{0}},m_{h_{t}});&&
C_{ij}^{j}=C_{ij}(-p_{t},p_{V},m_{t},m_{h_{t}},M_{Z});\\
C_{ij}^{k}=C_{ij}(-p_{t},p_{V},m_{t},M_{Z},m_{h_{t}}).&&
\end{eqnarray}
Here $g_{s}=\sqrt{4\pi\alpha_{s}}$, $T^{\alpha}$ are the Gell-Mann
$SU(3)_{c}$ matrices. $B_{i}$, $C_{ij}$ are two-point and
there-point scalar integrals. $p_{V}$ represents the momenta of
$Z$, $\gamma$, $g$, respectively.


\begin{thebibliography}{99}

\bibitem{y1}G. Eilan, J. L. Hewett and A. Soni, {\em Phys. Rev. D}{\bf 44},
1473(1991).
\bibitem{y2}R. Frey, {\em et al.}, FERMIAB-CONF-97-085,
hep-ph/9704243; J. A. Aguilar-Saavedra, G. C. Branco, {\em Phys.
Lett. B}{\bf 495}, 347(2000); J. A. Aguilar-Saavedra,
hep-ph/0012305.
\bibitem{y3}J. L. Diaz-Gruz, {\em et al.}, {\em Phys. Rev. D}{\bf 41},
891(1990).
\bibitem{y4}B. Grzadfkowski, J. F. Gunion and P. Krawcyzk, {\em Phys.
Lett. B}{\bf 268}, 106(1991); W. S. Hou {\em Phys. Lett. B}{\bf
296}, 179(1992).
\bibitem{y5}R. A. Diaz, R. Martine and J. Alexis Rodriguez, hep-ph/0103307.
\bibitem{y6}X. L. Wang, {\em et al}, {\em Phys. Rev. D}{\bf 50},
5781(1994).
\bibitem{y7}C. S. Li, R. J. Oakes and J. M. Yang, {\em Phys. Rev. D}{\bf 49}, 293(1994).
\bibitem{y8}C. S. Li, X. Zhang, S. H. Zhu, {\em Phys. Rev. D}{\bf 60}, 077702(1999)
\bibitem{y9}J. J. Cao, Z. Z. Hua and J. M. Yang,  hep-ph/0208033.
\bibitem{y10}C. T. Hill, {\em Phys. Lett. B}{\bf 345}, 483(1995); K. Lane, E. Eichten, {\em Phys. Lett. B}{\bf 352},
382(1995); K. Lane, {\em Phys. Lett. B}{\bf 433}, 96(1998).
\bibitem{y11}C. X. Yue, G. R. Lu, G. L. Liu, Q. J. Xu, {\em Phys. Rev. D}{\bf 64},
095004(2001).
\bibitem{y12}Hong-Jian He and C. P. Yuan, {\em Phys. Rev. Lett.
} {\bf 83},28(1999); G. Burdman, {\em Phys. Rev. Lett}{\bf
83},2888(1999).
\bibitem{y13}X. L. Wang, Y. L. Yang, B. Z. Li, C. X. Yue, J, Y, Zhang {\em Phys. Rev. D}
{\bf 66}, 075009(2000); X. L. Wang, Y. L. Yang, B. Z. Li, L.D.Wan,
{\em Phys. Rev. D} {\bf 66}, 075013(2000).J. J. Cao, Z. Z. Hua and
J. M. Yang, hep-ph/0212114.
\bibitem{y14}C. X. Yue, Y. P. Kuang, X. L. Wang, W. B. li, {\em Phys. Rev. D}{\bf 62},
055005(2000); C. T. Hill, E. H. Simmons, hep-ph/0203079.
\bibitem{y15}R. S. Chivukula, B. Dobrescu, H. Georgi and C. T.
Hill, {\em Phys. Rev. D}{\bf 59}, 075003(1999)
\bibitem{y16}J. Guasch, J. Sola, hep-ph/9906268.

\end{thebibliography}
 \end{document}